\documentclass[preprint,authoryear,12pt]{elsarticle}

\usepackage{graphics}

\usepackage{amssymb}

\journal{Journal of Atmospheric and Solar-Terrestrial Physics}

\begin{document}

\begin{frontmatter}



\title{Waveguide gravity disturbances in vertically inhomogeneous dissipative atmosphere}

\author[label1]{G.V. Rudenko}
 \ead{rud@iszf.irk.ru}
\author[label2]{I.S. Dmitrienko}
\ead{dmitrien@iszf.irk.ru}
\address[label1]{Institute of Solar-Terrestrial Physics SB RAS, Irkutsk, Russia}
\address[label2]{Institute of Solar-Terrestrial Physics SB RAS, Irkutsk, Russia}

\begin{abstract}
Trapped atmosphere waves, such as IGW waveguide modes and Lamb
modes, are described using dissipative solution above source
(DSAS) \citep{Dmitrienko2016}. Accordingly this description, the
modes are disturbances penetrating  without limit in the upper
atmosphere and dissipating their energy throughout the atmosphere;
leakage from a trapping region to the upper atmosphere is taken in
consideration. The DSAS results are compared to those based on
both accurate and WKB approximated dissipationless equations. It
is shown that the spatial and frequency characteristics of modes
in the upper atmosphere calculated by any of the methods are close
to each other and are in good agreement with the observed
characteristics of traveling ionospheric disturbances.

\end{abstract}


\end{frontmatter}


\section{Introduction}\label{section1}
This paper is devoted to describing  atmospheric trapped modes
extending to large heights. A description of these modes at higher
altitudes, above all, is extremely important in terms of their
experimental identification. The energy of the waveguide modes is
mainly concentrated in low heights, into their trapping region.
However, if the disturbance amplitude at these heights is very
small compared with the background atmospheric parameters, then we
have a possibility of indirect observation  the disturbance  in
the upper atmosphere only, where, due to an increase of relative
disturbance amplitude because of  the atmospheric density drop, it
can lead to significant changes in the charged component of the
ionosphere. It is because of the "invisible" IGW propagation in
the lower atmosphere we can observe such a phenomenon as the
traveling  ionospheric disturbances (TID), \citep{Hines1960}.

The problem of describing  the atmospheric trapped modes can be
solved, using  a dissipative solution above source (DSAS)
\citep{Dmitrienko2016}. Actually, any DSAS satisfies the upper
boundary condition for trapped modes, that is, the absence of the
flow of energy towards the Earth in the upper layers of the
atmosphere, and, because there is no source, is applicable up to
the Earth. Therefore, the problem of trapped mode finding reduces
to the problem of selection of a DSAS that satisfies  the lower
boundary condition for trapped modes - the zero displacement  on
the Earth's surface. Thus, the condition of the equality to the
zero of vertical velocity for DSAS on the Earth's surface is the
dispersion equation for trapped mode. This dispersion equation can
be regarded as an equation for a horizontal wavenumber at known
frequency or vice versa. We consider  IGW waveguide modes,  which
exist due to the temperature inhomogeneity of  the lower
atmosphere, and Lamb modes. Given a real frequency, we find the
complex horizontal wavenumber . The complexity of horizontal
wavenumbers is a consequence of dissipation  and presence of
subbarrier leakage from the IGW waveguide.  The horizontal
wavenumber found, we can obtain a vertical structure of the mode,
which  gives relationships of the disturbance parameters in the
lower atmosphere with its parameters in the upper isothermal
atmosphere.

We organize our paper as follows. A model of the atmosphere
applied for calculations is described in Section 2. Section 3 is
devoted to construction and analysis of waveguide modes of the IGW
spectral range. We compare the solutions of the waveguide problem,
constructed from a DSAS, to solutions obtained in the frames of
the dissipationless approximation by both the WKB method and
numerical ones.  Such comparisons aim at two things at once.
First, they allow us to reach clarity in understanding of
dissipation effect on the main characteristics of the waveguide
propagation: the dispersion relations, the waveguide leakage, and
the horizontal attenuation of the waveguide modes. Second, they
serve as additional  tests to the tests of \cite{Dmitrienko2016},
of both method for  DSAS and corresponding codes. We obtain
dispersive properties and a description of a height structure of
all disturbance parameters. All special features of trapped IGWs
are present in the obtained waveguide solution in the real
atmosphere: localization due to the temperature stratification;
the leakage through the opacity area; qualitative changes of the
wave structure related to dissipative nature of the disturbances
in the upper atmosphere. We obtain complete information on all the
height structure of waveguide modes, which can be directly used to
reveal a quantitative compliance of IGW modes with TIDs. We  show
that  the properties  of the obtained waveguide solutions are in
good agreement with the main characteristics of TIDs following
from their observations: the periods to space scales ratios, the
weak horizontal attenuation, the values of full phase velocity;
the inclinations of the phase fronts.

It should be noted that the waveguide modes were investigated in
\cite{Francis1973_a,Francis1973_b} and their results are widely
used both in theoretical works and in the interpretation of
observations of various disturbances, including those in the upper
atmosphere
\citep{Shibata1983,Afraimovich2001,Vadas2009,Vadas2012,Idrus2013,Heale2014,Hedlin2014}
In papers \cite{Francis1973_a,Francis1973_b}, the dispersion
characteristics and vertical structures of the waveguide modes
were obtained. In \cite{Francis1973__b}, it was shown that one or
two lower IGW modes are able present, due to the waveguide
leakage, at ionospheric heights. Method of Francis allows to
calculate well enough  the structure of wave disturbances in the
lower part of the atmosphere and the dispersion characteristics of
the modes captured  by the temperature inhomogeneity  of the lower
atmosphere. In details, method of Francis is discussed in terms of
its applicability in the upper atmosphere in
\cite{Dmitrienko2016}. Here we note only the fact that the
specificity of the method of Francis to use everywhere reduction
of order of the differential equations, allowable only in the weak
dissipation case, really do not give a possibility to obtain a
correct description of disturbances in the upper atmosphere. In
contrast to the method of Francis, in our method of constructing
of DRNI, we use reduction of order of the wave equation to the
second  (in our own way) for the small dissipation altitudes only,
where it is justified. By virtue of this, our method, unlike the
method of Francis, allows to describe disturbances of the upper
atmosphere adequately.

The calculations for the Lamb waves, analogous to those for the
IGW waveguide modes, are presented in  Section 4.  Section 5 is a
conclusion.
\section{Model of the atmosphere}\label{section2}
We make our calculations with usage  a model of the atmosphere
specified by the height profile of undisturbed temperature
$T_0(z)$ according to the NRLMSISE-2000 distribution with
geographic coordinates of Irkutsk for the local noon of winter
opposition: \\
 $p_0(z)=p_0(0)\exp\left[-\frac{g}{R}\int_0^z\frac{1}{T(z')}dz'\right], \ \ \ p_0(0)=1.01 \rm{Pa}$; \\
  $\rho_0(z)=\rho_0(0)\exp\left[-\frac{g}{R}\int_0^z\frac{1}{T(z')}dz'\right], \ \ \ \rho_0(0)=287.0 \rm{g/m^3}$. \\
Here  $p_0$, $\rho_0$   are the undisturbed density and pressure;
$z$, the vertical coordinate counted from the Earth's surface;
$g=9.807\rm{m/s^2}$, the free fall acceleration;
$R=287\rm{J/(kg\cdot K)}$, the universal gas constant. We take
into consideration that the atmosphere possesses thermal
conductivity, assuming its dynamic coefficient constant.

\begin{figure}
    \includegraphics[width=1\linewidth]{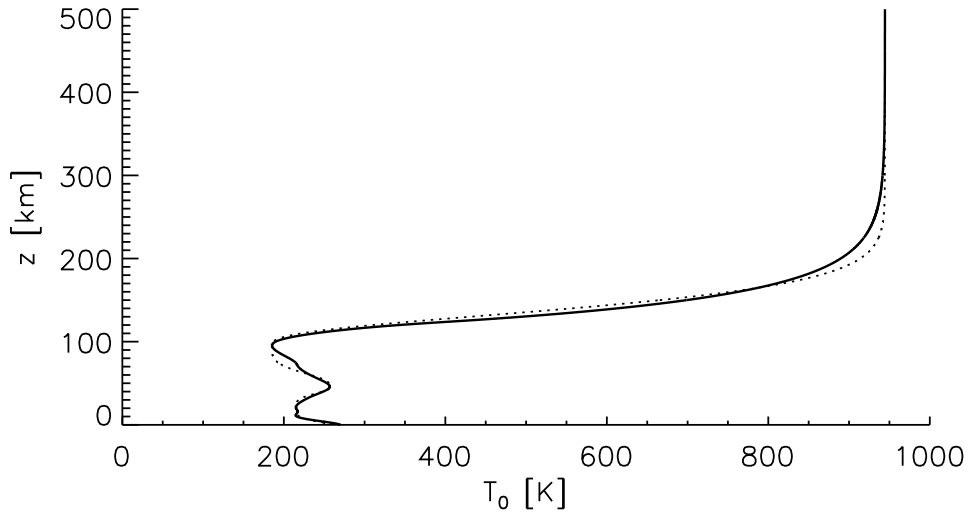}\\
  \caption{The approximate temperature dependence on height (dotted
line); the basic dependence (solid line).}\label{fig1}
\end{figure}

To make our calculations more precise, instead of an original
discrete function of the height-temperature distribution
$T_0(z)$, we have used its smooth approximation with a continuous
up to the third derivative function:
   \begin{equation}\label{eq62}
\begin{array}{l}
T(z>430 )=944.4, \\
T(95.3<z\leq 430 )=
\\
\left(\left[\cos\left(\frac{\pi}{2}\left(\frac{430-z}{430-95.3}\right)^6\right)\right]^3-1\right)(944.4-185.4)+944.4,\\
\\
T(46<z\leq 95.3 )=\\
-\left(\left[\cos\left(\frac{\pi}{2}\left(\frac{95.3-z}{95.3-46}\right)^2\right)\right]^3-1\right)(257-185.4)+185.4,\\
\\
T(20<z\leq 46 )=\\
\left(\left[\cos\left(\frac{\pi}{2}\left(\frac{z-20}{46-20}\right)^2\right)\right]^3-1\right)(215.1-257)+215.1,\\
\\
T(0<z\leq 20 )=\\
\left(\left[\cos\left(\left(\frac{20-z}{20}\right)^2\arccos\left(0.5^{3/2}\right)\right)\right]^3-1\right)(215.1-270.1)+215.1.
\end{array}
\end{equation}
Both the original dependence $T_0(z)$  and its approximation are
shown in Figure \ref{fig1}.
\section{IGW waveguide}\label{section3}
This section addresses long-period disturbances which occur at
ionospheric heights far from their sources. Seeing that such waves
cannot be trapped in the upper atmosphere (approximately
isothermal), the only way to explain the observations is to
consider the waves as a result of leakage from a waveguide located
at lower heights.

We compare three solutions of the waveguide problem: a) the WKB
approximation without dissipation, b) numerical solution of the
boundary value problem without dissipation, c) solution of the
waveguide problem using the DSAS \citep{Dmitrienko2016}.
Primarily, we will give necessary formulas for each of the
methods.
\subsection{Equations for the waveguide
problem}\label{section3.1}
\subsubsection{The WKB approximation for the waveguide modes without dissipation}\label{section3.1.1}
From the system of equations (3), (12), (13) in
\cite{Dmitrienko2016} in the principal order of the WKB
approximation, it is easy to get the square of the $k_z$
wavenumber as  $(k_z^{WKB})^2=U(z)$ where
\begin{equation}\label{eq63}
U(z)=-\frac{1}{4}\frac{\gamma^2g^2}{c^4_s}+(\gamma-1)\frac{k^2g^2}{c_s^2\omega^2}+\frac{\omega^2}{c_s^2}-k_x^2.
\end{equation}
We use the same notations as in \cite{Dmitrienko2016} in this
paper.

\begin{figure}
    \includegraphics[width=1\linewidth]{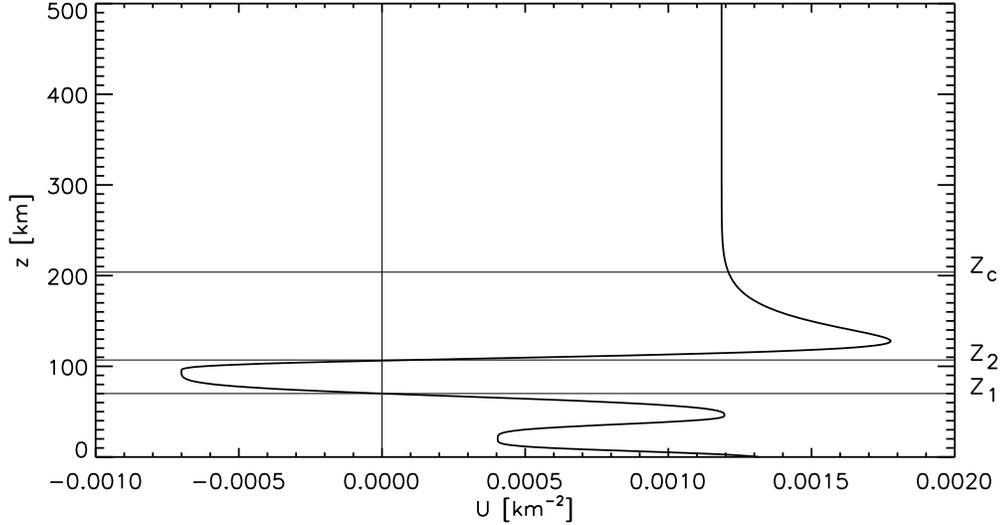}\\
  \caption{The characteristic height distribution of $U$-function.}\label{fig2}
\end{figure}

Discuss the  $U$-function profile (Figure \ref{fig2}) for
arbitrarily selected wave parameters $\omega$  and $k_x$
$\left(T_w=\frac{2\pi}{\omega}=90
min;\lambda_{hor}=\frac{2\pi}{k_x}=1390 km\right)$. It is evident
that the waveguide may be located below  $z_1$. The upper locking
barrier of the waveguide is a region of negative  $U$-values:
$z_1<z<z_2$. Above $z_2$  is again a region of wave propagation.
Since the barrier at $z_1<z<z_2$  has a finite thickness, the
waves  can leak partially  from the region $z<z_1$ to the region
$z>z_2$.

A distinguishing characteristic of the problem in hand is a strong
variation of  $U$  form and values depending on wave parameters
$\omega$ and $k_x$. The values of $z_1$  and $z_2$  change too.

In case of subbarrier leakage, one can show that the waveguide
capture condition may be written in form of a modified
Bohr-Sommerfeld condition of quantization with  complex turning
points:
         \begin{equation}\label{eq64}
         \begin{array}{l}
\int_C\sqrt{U(z)}dz\approx\pi\left(\frac{1}{2}+n\right)+i\exp\left[-2\int_{z_1}^{z_2}\sqrt{|U_0(z)|dz}\right]\equiv\\
\pi\left(\frac{1}{2}+n\right)+iQ, \ n=0,1,... \ .
\end{array}
\end{equation}
The condition of Eq. (\ref{eq64}) gives dispersion relations
between real frequencies $\omega$  and complex wave numbers, $k_x$
with a small imaginary part accounting for horizontal attenuation
of the waveguide mode $n$. The integration contour  $C$ of the
integral on the left-hand side of Eq. (\ref{eq64}) begins from
$z_0$ and ends at a complex turning point $z_{c1}$ close to the
real turning point  $z_1(U(z_1,\omega,{\rm Re}k_x))$. Besides,
with inner (complex) turning points, we assume that the
$C$-contour also passes through these points (in our calculations,
for simplicity, sections of the $C$-contour with ${\rm Re}U(z)<0$
are ignored.) The integral in the exponent argument on the
right-hand side of Eq. (\ref{eq64}) is assumed to be real; the
$_0$-index of the $U$-function in the integrand means that it is a
function of the real part $k_x$  and real $z$:
$U_0(z)=U(z,\omega,{\rm Re}k_x)$. To solve Eq. (\ref{eq64}) we use
the perturbation theory. From the equation
          \begin{equation}\label{eq65}
\int_{0}^{z_1}\sqrt{|U_0(z)|dz}-\pi\left(\frac{1}{2}+n\right)=0
\end{equation}
for the selected real   $\omega$, we find a real root  $k_{0x}$ in
Eq. (\ref{eq65}). Next, we write
          \begin{equation}\label{eq66}
k_x=k_{0x}(1+i\delta), \ U(z)=U_0(z)+k_{0x}i\delta\frac{\partial
}{\partial k_x}U_0;
\end{equation}
By substituting Eq. (\ref{eq66}) to Eq. (\ref{eq64}) and
accounting for the complexity of turning points $z_{cj}$ of the
integration contour of the integral on the left-hand side of Eq.
(\ref{eq64}), we obtain an equation for the complex addition of
the horizontal wave vector:
          \begin{equation}\label{eq67}
\delta\frac{1}{2}\int_0^{z_1}\frac{k_{0x}}{\sqrt{U_0}}\left(\frac{\partial}{\partial
k_x}U_0\right)dz+\delta^{3/2}\frac{2}{3}e^{3i\pi/4}k_{0x}^{3/2}\sum\limits_{j}\left[\frac{(\partial
 U_0/\partial k_x)^{3/2}}{|\partial U_0/\partial z|}\right]_{z=z_j}=Q.
\end{equation}
Eq (\ref{eq67})  has three roots  $\delta^{3/2}$. We choose one of
them that corresponds to attenuation of the waveguide mode.
\subsubsection{Boundary value problem (BVP) for waveguide modes without dissipation}\label{section3.1.2}
It is most convenient to formulate the boundary value problem on
the base of a second order differential equation for disturbed
vertical velocity   $v_z$.  It is easy to get such an equation
from a set of equations (3), (12), (13) in \cite{Dmitrienko2016}:
\begin{equation}\label{eq13}
\begin{array}{l}
 p'=a_{11}p+a_{12}v_z,\\
v'_z=a_{21}p+a_{22}v_z,
  \end{array}
\end{equation}
where: \\
  $a_{11}=\frac{p'_0}{\gamma p_0};
  a_{12}=i\omega\rho_0\left(1-\frac{\omega_N^2}{\omega^2}\right);$\\
$a_{21}=-\frac{i\omega}{\rho_0}\left(\frac{p'_0}{\gamma
p_0g}+\frac{k_x^2}{\omega^2}\right);a_{22}=-\frac{p'_0}{\gamma
p_0}.$  \\
The second equation is to be differentiated from (\ref{eq13}):
         \begin{equation}\label{eq14'}
v''_z=a_{21}p'+a_{22}v'_z+a'_{21}p+a'_{22}v.
\end{equation}
Then using $p'$  for the first equation from (\ref{eq13}) and
expressing $p$ from the 2nd equation from (\ref{eq13}):
$p=\frac{1}{a_{21}}(v'_z-a_{22}v_z)$,we obtain an equation for
$v_z$
\begin{equation}\label{eq14''}
\begin{array}{l}
 v''_z=Pv_z+B=0,\\
P=-a_{11}-a_{22}-({\rm
ln}a_{21})'=\frac{T'_0}{T_0\left(\frac{\omega^2}{k^2_xc_s^2}-1\right)}-\frac{\gamma
g}{c_s^2},\\
B=-a_{21}a_{12}+a_{11}a_{22}+a_{22}({\rm ln}a_{21})'-a'_{22}=\\
\ \ \
=-\frac{g}{c_s^2}P-\left(\frac{g}{c_s^2}-\frac{T'_0}{T_0}\right)\frac{g}{c_s^2}+\frac{g}{c_s^2}\left(\frac{\omega^2}{c_s^2}-k_x^2\right)\left(1-\frac{\omega_N^2}{\omega^2}\right).
  \end{array}
\end{equation}
It is most convenient to solve the boundary value problem for Eq.
(\ref{eq14''}), using the corresponding nonlinear Riccati
equation:
          \begin{equation}\label{eq68}
G'-PG-BG^2-1=0,
\end{equation}
where: $G$  is related to  $v_z$ through
          \begin{equation}\label{eq69}
Gv'_z=v_z.
\end{equation}
The  $G$-function of the waveguide solution should meet the upper
and lower boundary conditions. At the top ($z=z_\infty\to\infty$),
the $G$-function should fit an upward IGW:
          \begin{equation}\label{eq70}
G(z_\infty)=-\left(\frac{1}{2}P_\infty+\sqrt{B_\infty-\frac{1}{4}P_\infty}\right)^{-1}.
\end{equation}
In the numerical implementation,  $z_\infty$ was taken to be
$430km$, above which, in our model, the  $U$-function is constant.
At the bottom ($z=0$), we impose a requirement:
\begin{equation}\label{eq71}
G(0)=0.
\end{equation}
This requirement, according to Eq. (\ref{eq69}), is equivalent to
the condition of equality to the zero of the  $v_z$.

The numerical solution of a Cauchy problem Eqs. (\ref{eq68}),
(\ref{eq70}) in the $G$-function allows us to obtain a complex
dispersion equation
\begin{equation}\label{eq72}
D(\omega,k_x)=G(0,\omega,k_x).
\end{equation}
Formally, we can solve Eq. (\ref{eq72}) by assuming the first or
second argument of dispersion  $D$-function to be real. In the
former case, we will have modes attenuating in the horizontal
direction of propagation; in the latter case, modes attenuating in
time. In this paper, we analyze the modes only with real values of
the frequency  $\omega$. A vertical spatial structure of the mode
(for the pair of  values  $\omega$ and $k_x$ satisfying Eq.
(\ref{eq72})) can be obtained by solving numerically a Cauchy
problem for Eq. (\ref{eq14''}) with the initial condition
$v_z(0)=0$, $v'_z(0)=1$ corresponding to the boundary condition of
Eq. (\ref{eq71}) for the  $G$-function.
\subsubsection{Solution of the waveguide problem using the DSAS}\label{section3.1.3}
Because the upper boundary conditions for waveguide modes are the
same as for the DSAS \citep{Dmitrienko2016}, it is enough to
choose the DSAS satisfying the condition of the equality to the
zero of vertical speed on the Earth's surface for the solution of
the waveguide problem. Thus, the dispersion equation becomes:
\begin{equation}\label{eq73}
D(\omega,k_x)=v_z(0,\omega,k_x).
\end{equation}
It is complex following  subbarrier leakage and presence of
dissipation.
\subsection{Numerical calculations of dispersion dependencies of internal gravity waveguide modes}\label{section3.2}
\begin{figure}
    \includegraphics[width=1\linewidth]{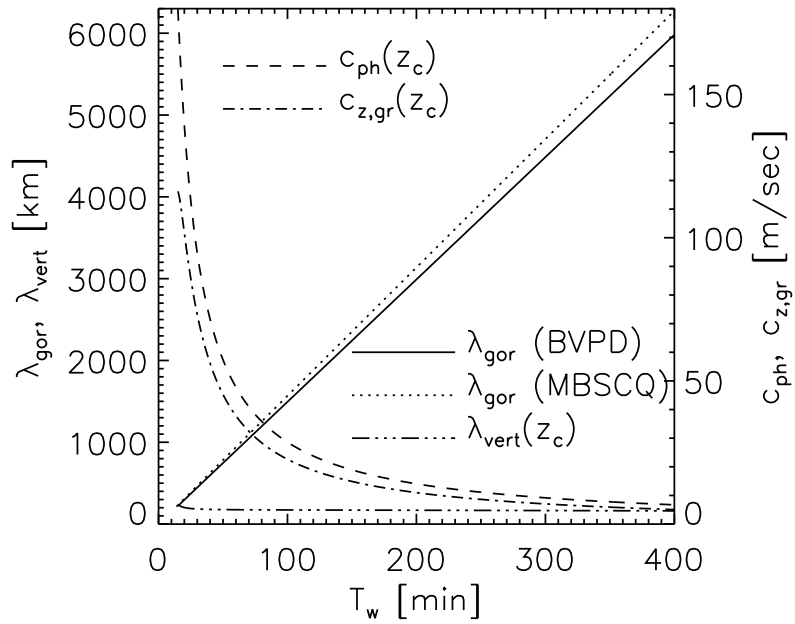}\\
  \caption{The waveguide characteristics of  -mode: horizontal wavelength from the BVPD = BVP (solid line);
  horizontal wavelength from the MBSCQ (dotted line); full phase velocity of the upward propagating wave
  (dashed line); vertical group velocity of the upward propagating wave (dash-dotted line); vertical length
  of the upward propagating wave (dash-dot-dotted line).}\label{fig10}
\end{figure}

\begin{figure}
    \includegraphics[width=1\linewidth]{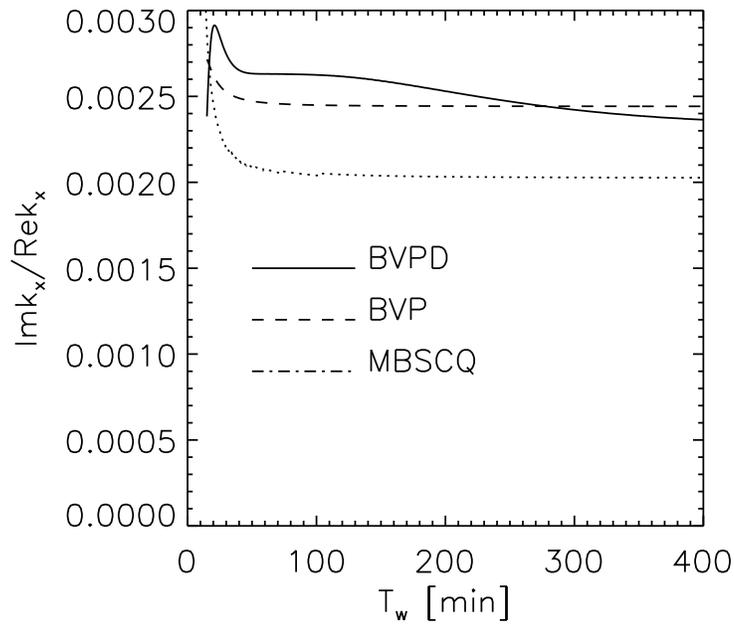}\\
  \caption{The waveguide characteristics of 0-mode: horizontal attenuation
  characteristic from the BVPD (solid line); horizontal attenuation characteristic
  from the BVP (dashed line); horizontal attenuation characteristic from the MBSCQ (dotted line).}\label{fig11}
\end{figure}

First, we have established that there is only one nodeless
waveguide mode (with $n=0$) in the selected model of the
atmosphere in the frequency range that corresponds to TIDs. This
was obtained by all the algorithms described in Section 3.1. For a
more detailed analysis of propagation characteristics of the
waveguide mode, Figure \ref{fig10} gives: \\
--    the dispersion dependence of horizontal wavelength on
oscillation period (the solid curve is the dependence obtained
from the BVPD = BVP; the dotted curve, from the MBSCQ); \\
--    the characteristics of the wave leakage to the upper
atmosphere for reference height  $z=z_c$: full phase velocity
(dashed curve); vertical group velocity (dash-dotted curve);
vertical wavelength (dash-dot-dotted curve). \\
Figure \ref{fig11} presents characteristics of the horizontal
attenuation: the solid curve is the characteristics obtained from
the BVPD; the dashed curve, from the BVP; the dotted curve, from
the MBSCQ.

Note the most important points: \\
--    The case of the model of the atmosphere, we analyzed, has
showed that there was only one mode. Seeing that the chosen time
of the model for the geographic localization considered
corresponds most often to the moments of detection of ionospheric
disturbances, we may assume that the implementation of the
conditions (somewhere) for two or more modes is most likely to be
extremely rare or impossible at all. \\
--   The three approaches (the BVPD, BVP and MBSCQ) show close
values for phase horizontal velocity and attenuation. The
attenuation is  sufficient  small for very long-distance
propagation of waveguide disturbances. The fact that the WKB
approximation,  despite its formal invalidity for  $0$-mode,
provides results which are close to the exact solution, gives the
base for its usage at least for estimation. \\
-- It is interesting to note a property which is likely to be
specific only for IGW waveguides. The dependence in Figure
\ref{fig11} shows a high quality  factor of oscillations relative
to the characteristic of horizontal attenuation in spite of the
fact that, in the opacity barrier  $[z_1,z_2]$, the amplitude of
the solution decreases slightly. The parameter  $\sqrt Q$
determined in Eq. (\ref{eq64}) is of order of  $0.41$. For an
acoustic waveguide, its horizontal attenuation characteristic
would be estimated as  $0.41^2$. For IGWs, the multiplier of the
first term in Eq. (\ref{eq67}) gives a value of order of $30$ (for
AGW, it would be $\sim 1$) on the whole dispersion curve. This
factor causes a very weak attenuation of a mode. From a physical
point of view, this effect is provided by the smallness of the
vertical group velocity of the leaking wave.

It is important that the obtained dispersion dependence fairly
faithfully reproduces the observed ratio of horizontal scales to
periods of TIDs and gives reasonable estimates for their full
phase velocities obtainable in measurements \citep{Ratovsky2008,
Medvedev2009} These results are quite consistent with the concept
of the waveguide nature of the TIDs. The more detailed comparisons
of waveguide-disturbance properties with the observed properties
of TIDs allow us to implement the latest results obtained in
\cite{Medvedev2013}. In this paper the spatio-temporal structure
of traveling ionospheric disturbances characteristics is studied
on the base of the electron density profiles measured by two beams
of the Irkutsk incoherent scatter radar and the Irkutsk Digisonde.
First of all, we will compare the values of \emph{Elevation} and
\emph{Velocity} obtained from the 12-h window spectral analysis
(shown in Fig.1, \cite{Medvedev2013}) with the angular
characteristic ${\rm Atan}\frac{k_{vert}(z_c)}{k_{hor}(z_c)}$ and
velocity $c_{ph}(z_c)$ accordingly equivalent to them (Table
\ref{table1}).
\begin{table}
\caption{The angular and velocity characteristics}
 \label{table1}
\begin{tabular}{ccccccc}
\hline
                                    & $T_w=82min$ & $T_w=182min$ & $T_w=82min$& $T_w=182min$ \\
                                    & \emph{Elevation}    & \emph{Elevation}       & \emph{Velocity} &
                                    \emph{Velocity}\\
\hline
Medvedev & & & \\
at. al. (2013)  & $[-78^o,-71^o]$              & $[-86^o,-73^o]$     & $[12,28](m/c)$   & [19.5,32](m/c)   \\
                                    &                     &                         &         &          \\
Our values & $-82^0$              & $-86^0$                  & 35 (m/c)   & 15.6 (m/c)   \\
 \hline
\end{tabular}
\end{table}
The values shown in the table above are a matter of judgement
because they contain $k_vert$  which,generally  speaking, is not
applicable  at heights of the order of  $z_c$. Besides that, the
presence of real wind at the considered heights not taken into
consideration in theory can lead to noticeable differences between
the observed characteristics and theoretical ones. The fact that
the characteristics predicted in theory correspond in their values
to the observed ranges of these characteristics is enough for us.
Comparison of the most probable values of \emph{Elevation},
\emph{Velocity} and \emph{Wavelength} $(\approx \lambda_{vert}  )$
given in Figures 4-6 from \cite{Medvedev2013} is also of interest.
In accordance with a diurnal model of the atmosphere used here, we
are interested in left-hand parts of these figures for daytime
recordings.    Note that the time signal spectrum in daytime
recordings corresponds to the limited interval of periods from
approximately  $\sim 1h$ to $\sim 3 h$  with two local probability
peaks in the neighborhood of periods  $3 h$ and $1.5 h$ (Figure 2,
\cite{Medvedev2013}). The first peak has a largest value and a
small width; the second, a smaller value, diffused. Our values of
\emph{Elevation} and \emph{Wavelength} demonstrate the best match
with the observations. The most probable value of
\emph{Elevation}, (Figure 4, \cite{Medvedev2013}) of $-75^o$  is
close to our values (see Table \ref{table1}). The most probable
value of \emph{Wavelength} (Figure 6, \cite{Medvedev2013}) of $175
km$ is very close to our value $\lambda_{vert} = 192 km$ (see
Figure \ref{fig10}). The most probable value of \emph{Velocity}
(Figure 5, \cite{Medvedev2013}) of $35 m/c$ corresponds to the
value of the period $T_w=1.4 h$ in our plot shown by Figure
\ref{fig10}. This value $T_w$ corresponds to one of the spectral
distribution maxima (Figure 2, \cite{Medvedev2013}) Thus, we see
that our theoretical description is well supported by the
observational facts.
\subsection{Height structure of  an IGW waveguide mode}\label{section3.3}
\begin{figure}
    \includegraphics[width=1\linewidth]{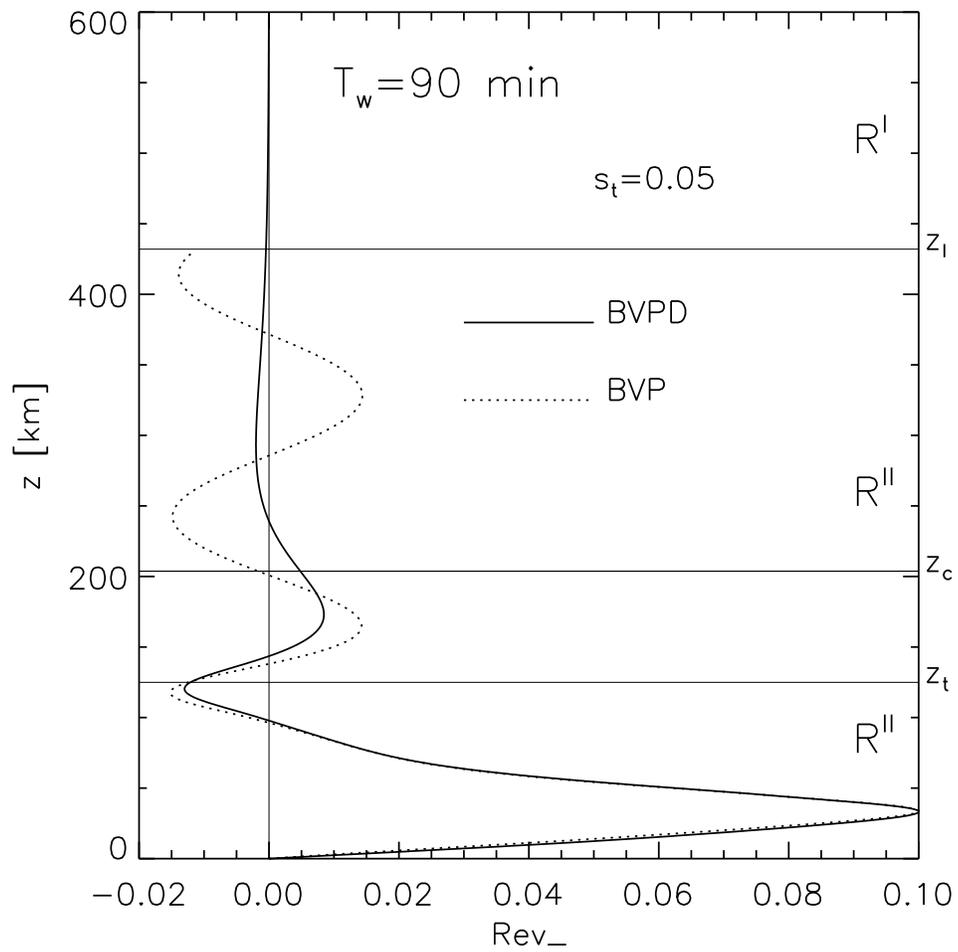}\\
  \caption{The example of comparison of a BVPD and BVP solutions.}\label{fig12}
\end{figure}

After the waveguide dispersion relations found, we need to
calculate the DSAS for two wave parameters $\omega$ and $k_x$
connected by the dispersion dependence to obtain the height
structure of a waveguide solution. As for BVPD method, this
procedure is contained already in it.  By the BVP method , we can
get dependence $v_z(z)$, using the function $G(z,\omega,k_x)$
proceeded from the BVP. It is sufficient for it to numerically
integrate a first-order differential equation $v'_z=v_z/G$  with
the initial value  $v_z(z_\infty)=1$. We will use a result of the
BVP waveguide solution only to compare it with a solution by our
principal BVPD method. For comparison, we use height distributions
$v_\_(z)=\left(\frac{\rho_0}{\rho_0(0)}\right)^{1/2}\frac{v_z}{c_s}$.
Figure \ref{fig12} shows ${\rm Re}v_\_$  for BVPD and BVP methods
when  the wave period is $T_w=90 min$ and the corresponding
dispersion
values of horizontal wavenumbers are: \\
$k_x=(4.66233\times 10^{-3}+i1.22515\times 10^{-5})km^{-1}$ (BVPD)\\
$k_x=(4.66215\times 10^{-3}+i1.14273\times 10^{-5})km^{-1}$ (BVP)\\
We see that both solutions are very close to each other in the
region $R^{III}$  where dissipation is small. Higher their plots
are naturally strongly different. The  BVP solution  tends to an
asymptotes of a simple wave in a homogeneous medium, and the DSAS
of the BVPD  attenuates under the effect of wave dissipation.
Thus, Figure \ref{fig12} clearly shows that a wave description
without dissipation is true only for the limited height range
where the smallness condition of the parameter $s$  is satisfied.

For ease of a graphic presentation of the BVPD solution, we will
use two sets of functions. For the top of a waveguide solution, we
use $\Theta,n,f,v,u$ determined by Eqs. (21) in
\cite{Dmitrienko2016}. For the bottom, we use their following
modifications:
\begin{equation}\label{eq74}
\begin{array}{l}
 \Theta^-=W\Theta,\\
n^-=Wn,\\
f^-=Wf,\\
v^-=10\times Wv,\\
u^-=Wu.
  \end{array}
\end{equation}
Where $W(z)=\left(\frac{\rho_0(z)}{\rho_0(200 km)}\right)^{1/2}$.
In all the next solutions presented, we use normalization: $
Max\left(\sqrt{v_x^2+v_z^2}\right)=50 m/c$

\begin{figure}
    \includegraphics[width=1\linewidth]{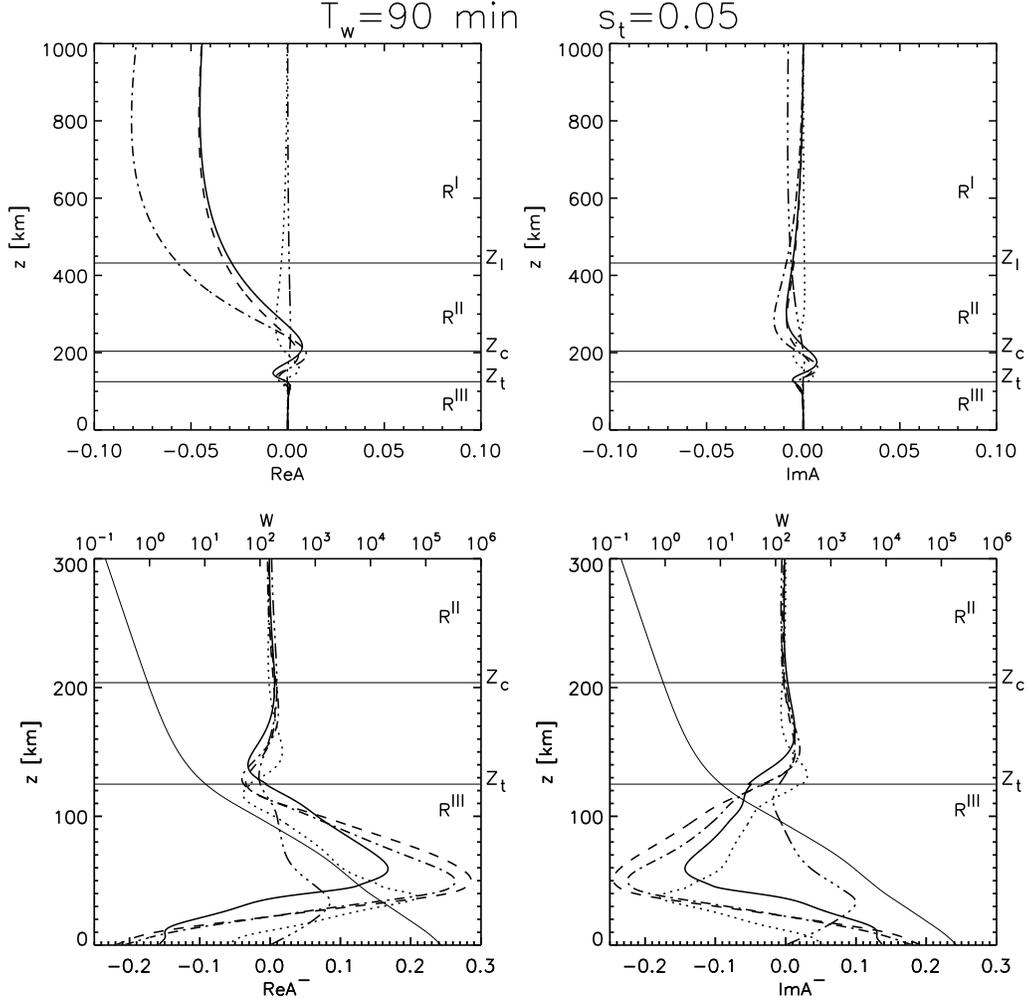}\\
  \caption{The full structure of a waveguide solution for the
  period $T_w=90 min$
  Match of lines with wave values is the same as in Figure 6 in \cite{Dmitrienko2016} (point -- $\Theta^(\Theta^-)$; dash-point-point-point --
  $v(v^-)$; solid -- $n(n^-)$; dashed -- $f(f^-)$; dash-point -- $u(u^-)$  ).
  The thin solid line describes the height distribution of the function  $W(z)$.}\label{fig13}
\end{figure}

Let us discuss the first example of a waveguide structure in
Figure \ref{fig13}.
Dependencies of real and imaginary parts of components of
disturbance of Eqs. (\ref{eq74}) and ((21) in
\cite{Dmitrienko2016}) are reflected in this figure. The top plots
show dependence of relative wave disturbances. The bottom plots
describe all the details of the waveguide structure which we
cannot see on the top plots because of the exponential factor. We
see that the relative values are the largest in the upper
atmosphere; the absolute ones, in the lower one. Reaching some
maximum, the relative values start decreasing due to effect of
dissipation. All the values look continuous at the height of $z_s$
with the presented scale. The solution discontinuity indicators
are sufficiently  small in this case:
\begin{equation}\label{eq75}
\begin{array}{l}
\delta_{[\Theta]}=\delta_{[T]},\delta_{[\Theta^-]}=0.134,\\
\delta_{[n}=\delta_{[\rho]},\delta_{[n^-]}=0.095.
  \end{array}
\end{equation}
Thus, we got a solution close in quality to the solutions for the
isothermal model of the atmosphere (see Section 5 in
\cite{Dmitrienko2016})  Note that the calculations show that the
discontinuity indicators are identical in order for all the
dispersion curve points.

\begin{figure}
    \includegraphics[width=1\linewidth]{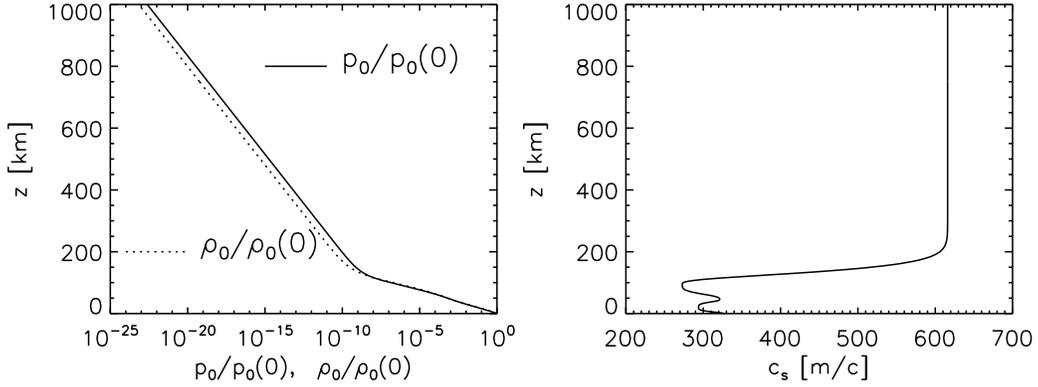}\\
  \caption{The height dependencies of undisturbed pressure, density, and sound velocity.}\label{fig13a0}
\end{figure}

The undisturbed temperature (Figure \ref{fig13}), the pressure,
the density, and the sound velocity (Figure \ref{fig13a0}) shown
in the bottom of Figure \ref{fig13}   (thin solid line) give the
possibility of calculating absolute values of the disturbances at
any height.

\begin{figure}
    \includegraphics[width=1\linewidth]{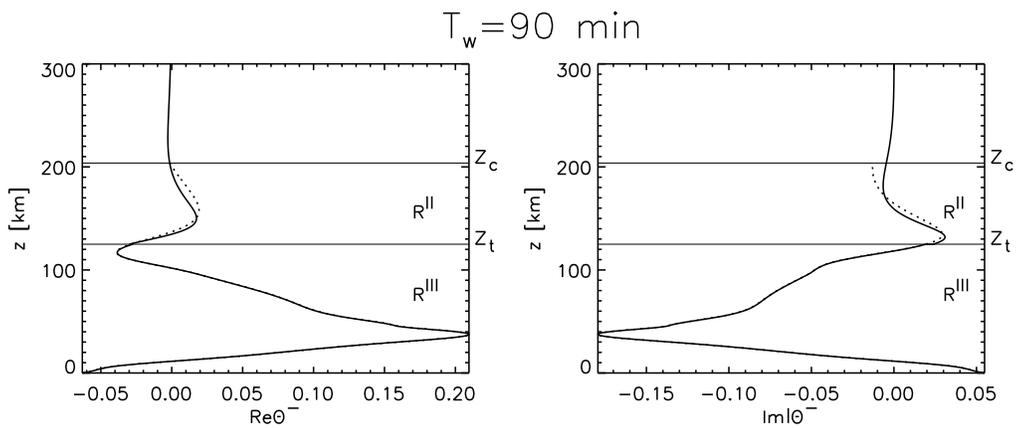}\\
  \caption{The comparison of the waveguide solutions with and without dissipation.}\label{fig13a}
\end{figure}

\begin{figure}
    \includegraphics[width=1\linewidth]{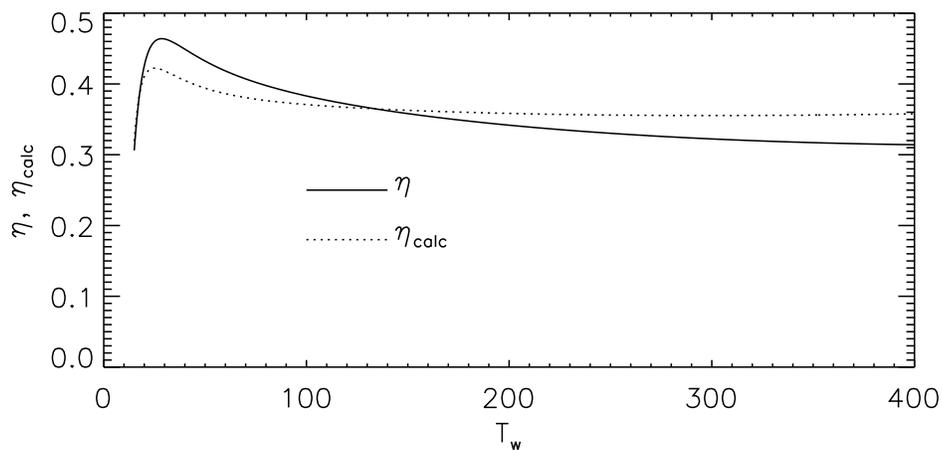}\\
  \caption{The comparison of $\eta_{calc}$  and  $\eta$.}\label{fig13b}
\end{figure}

For an indirect test, one can also use a calculation of a
numerical parameter similar to that of total vertical dissipation
index $\eta$   of (47)  in \cite{Dmitrienko2016}) for a isothermal
model of the atmosphere. In the isothermal case, $\eta$  is a
ratio of the amplitude modulus $T$  at the height of $z_c$  to the
amplitude modulus $T$ of upwards propagating  wave at the same
height without accounting for dissipation. If, as in our case, a
part of a reflected wave at the height of  $z_c$ is small, instead
of an incident wave without dissipation, we may use a full
combination of an incident wave with a reflected one.  Proceeding
from it, we select $\eta_{calc}$, for the case with an
inhomogeneous atmosphere, similar to $\eta$. To calculate this
value, we solve a Cauchy problem from the bottom to the top up to
the height $z_c$ for the system of equations (3), (12), (13) of
Case III from \cite{Dmitrienko2016}, using $p$  and $v_z$  of the
DSAS as boundary values on the Earth's surface. We calculate the
value $\eta_{calc}$ as a ratio of the modulus $T$ at the height of
$z_c$ to the modulus $T$ of the wave solution got at the same
height without dissipation. We give an example of such a
calculation in Figure \ref{fig13a}. This figure confirms once more
that the noticeable distinctions between the solutions with and
without dissipation begin with the height $z_s$. It can be seen
that these solutions significantly differ at the height of $z_c$.
We compare the values of $\eta_{calc}$  and  $\eta$ calculated in
the theory for the isothermal model where the backgraund
temperature is chosen of  $T_0$ in the non-isothermal model at the
height of  $z_c$. A comparison result of these values in the
$0$-mode spectrum is shown in Figure \ref{fig13b}.
The last figure shows order of the vertical total dissipation
index within the interval  $z=[-\infty,z_c]$;
$\eta(-\infty,z_c)\approx \eta(z_s,z_c)$. The fact that
$\eta_{calc}$ is close to isothermal $\eta$  is an essential
confirmation of the correctness of the DSAS. Note that if the
values $\eta_{calc}$ ,  $\eta$ are defined through the complex
values of their definiens, the complex $\eta_{calc}$ ,  $\eta$
coincide with each other roughly with the same precision. We do
not give the results of calculations for the complex $\eta_{calc}$
,  $\eta$.

\begin{figure}
    \includegraphics[width=1\linewidth]{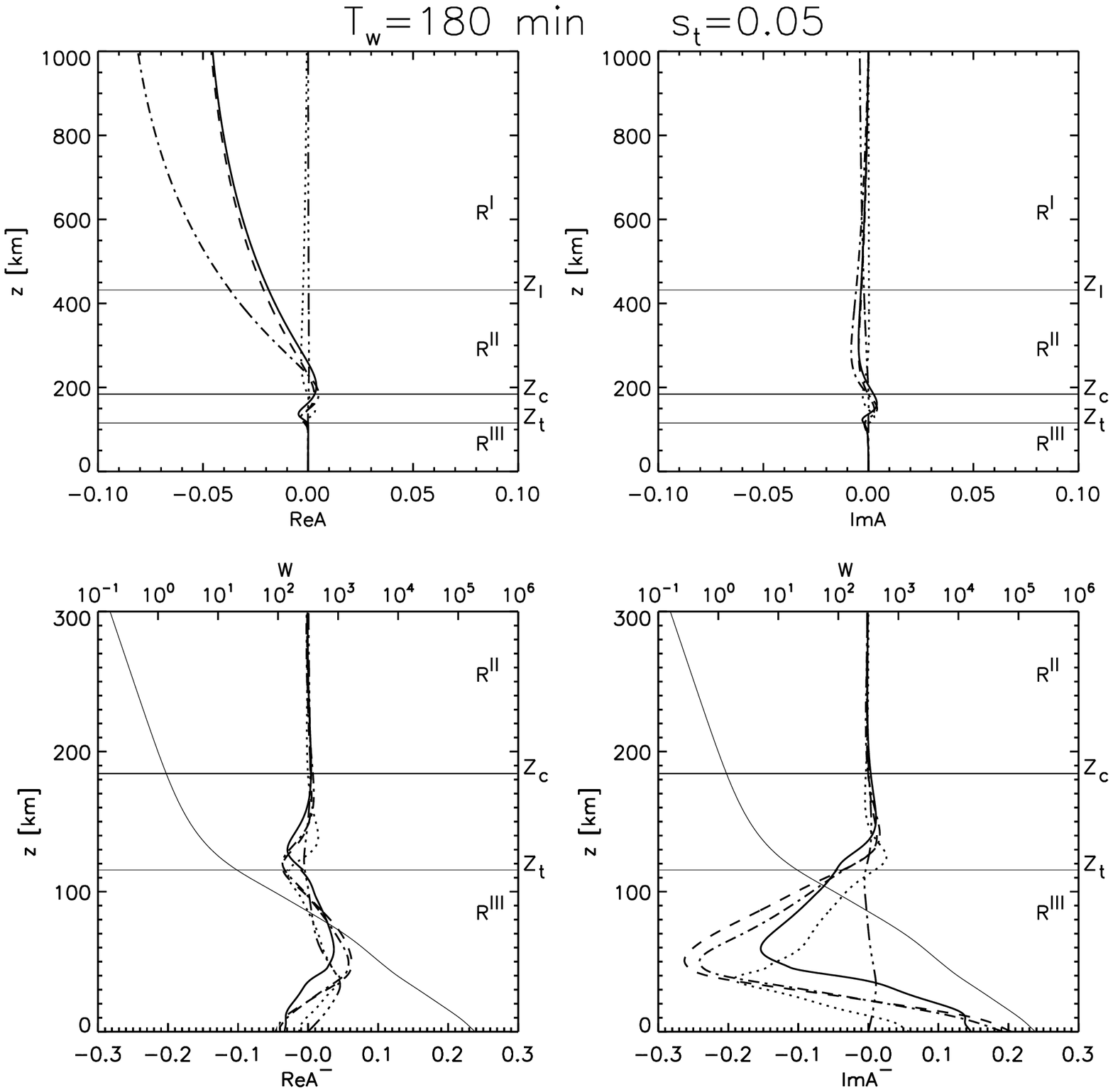}\\
  \caption{The complete structure of a waveguide solution for the period $T_w=180 min$.}\label{fig14}
\end{figure}

\begin{figure}
    \includegraphics[width=1\linewidth]{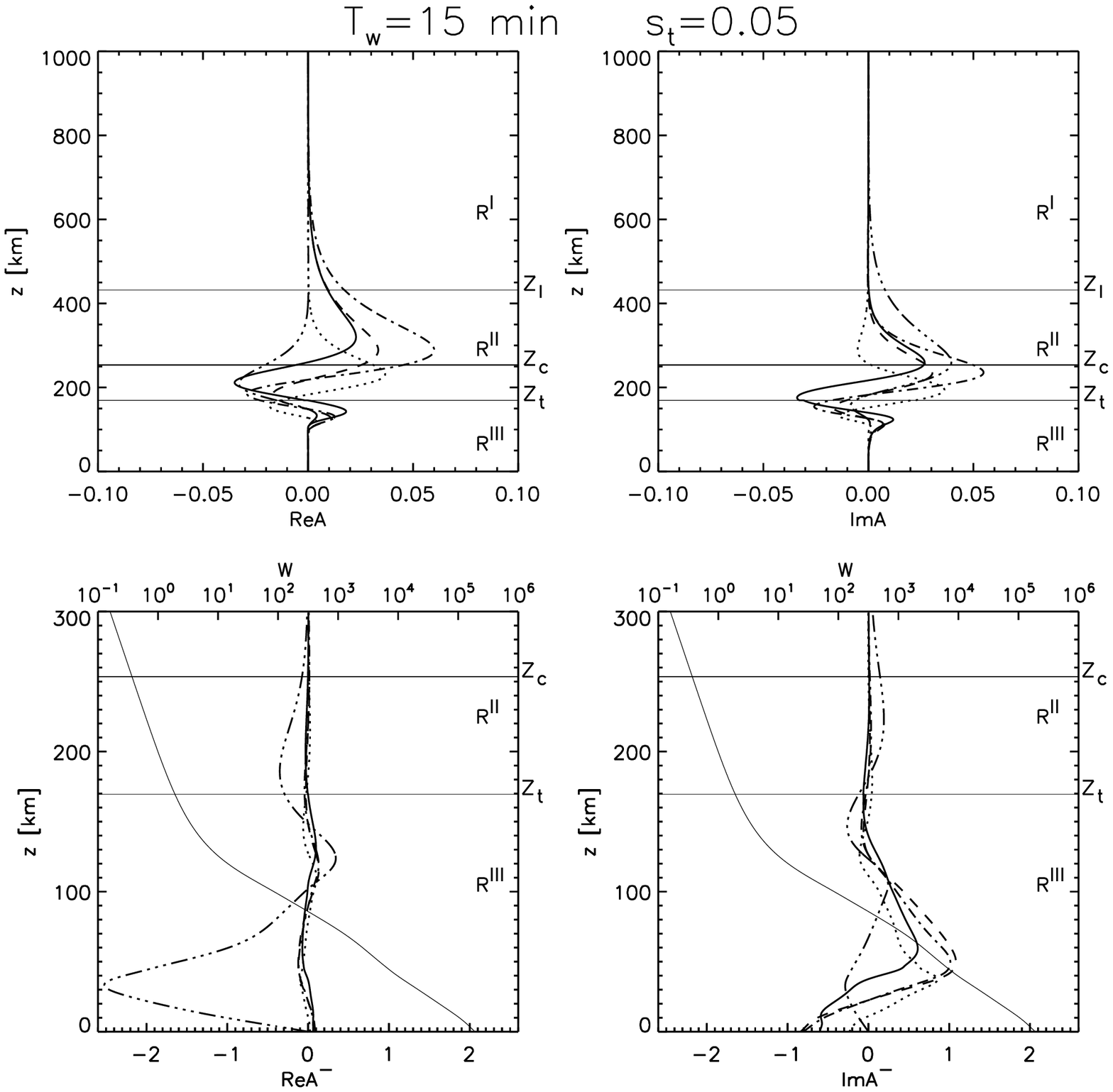}\\
  \caption{The complete structure of a waveguide solution for the period $T_w=15 min$.}\label{fig15}
\end{figure}

\begin{figure}
    \includegraphics[width=1\linewidth]{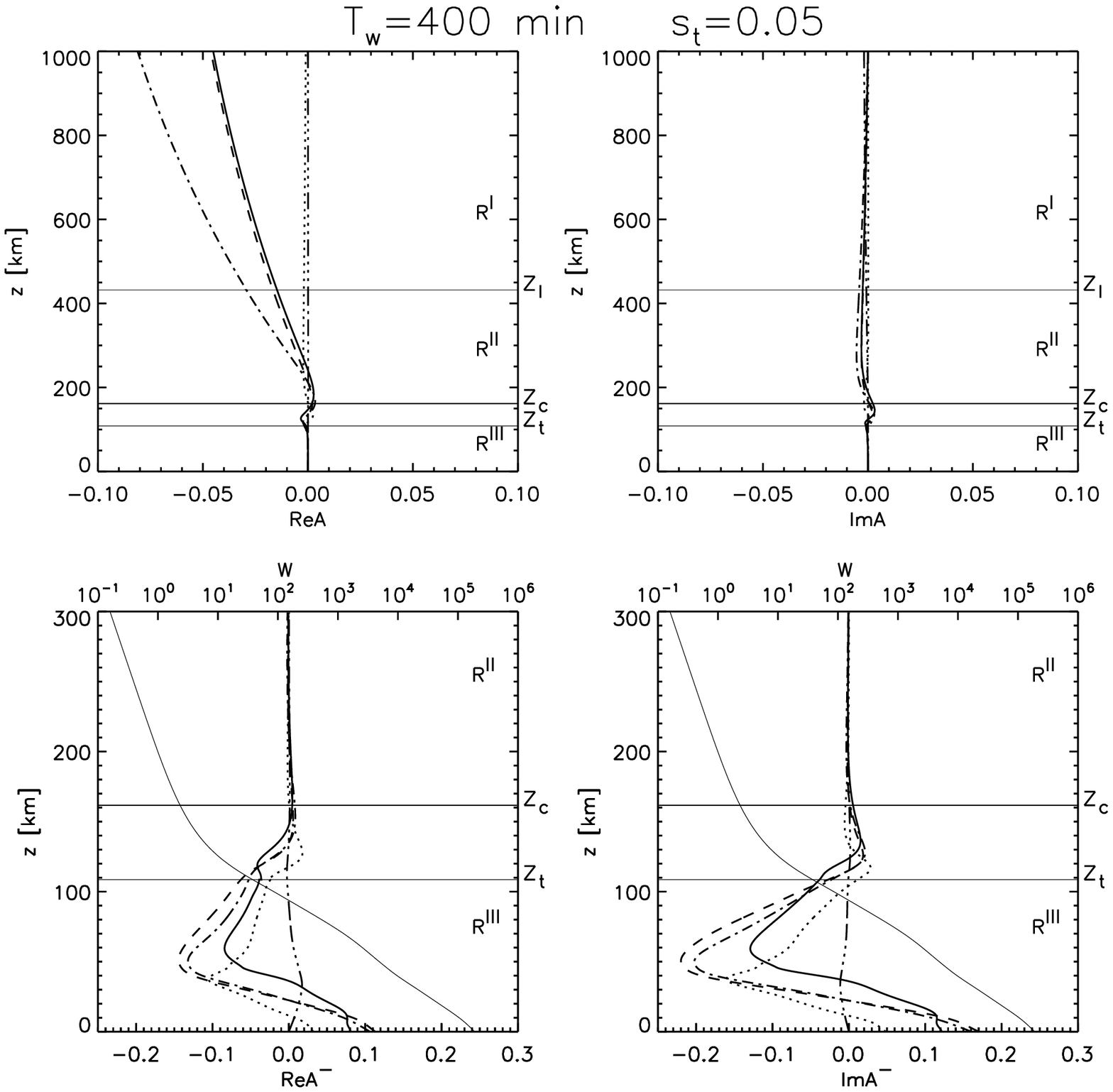}\\
  \caption{The complete structure of a waveguide solution for the period $T_w=400 min$.}\label{fig16}
\end{figure}

For a more complete understanding of a waveguide mode, we present
three more figures, Figures \ref{fig14}-\ref{fig16}, for $T_w=180
min$ and extreme points of the spectral range $T_w=15 min$  and
$T_w=400 min$.
Note that the longwave spectrum range (Figures \ref{fig14}
-\ref{fig16}) does not see discontinuation of the growth of some
relative values in the calculated height interval, but it is clear
that they attenuate beyond it because the evanescent analytical
solutions are continuations of the numerical solutions.
\section{Lamb waves}\label{section4}
\begin{figure}
    \includegraphics[width=1\linewidth]{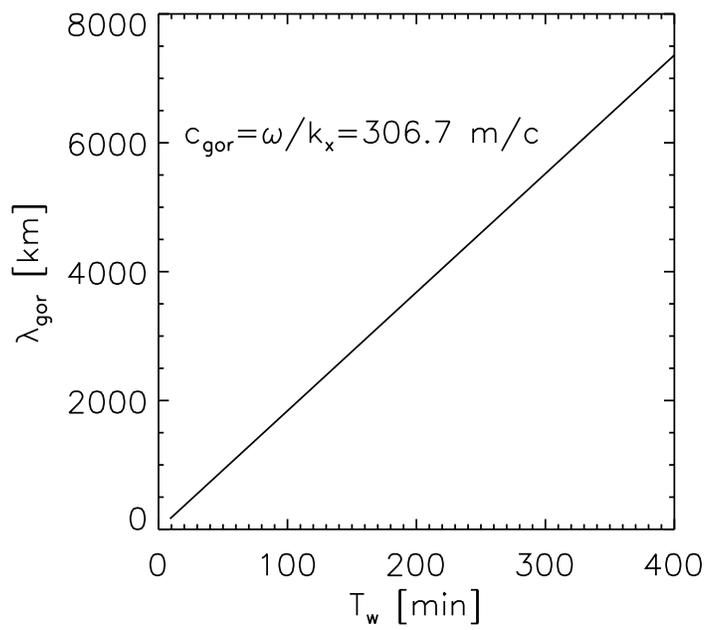}\\
  \caption{The dispersive relation for a Lamb mode.}\label{fig17}
\end{figure}

\begin{figure}
    \includegraphics[width=1\linewidth]{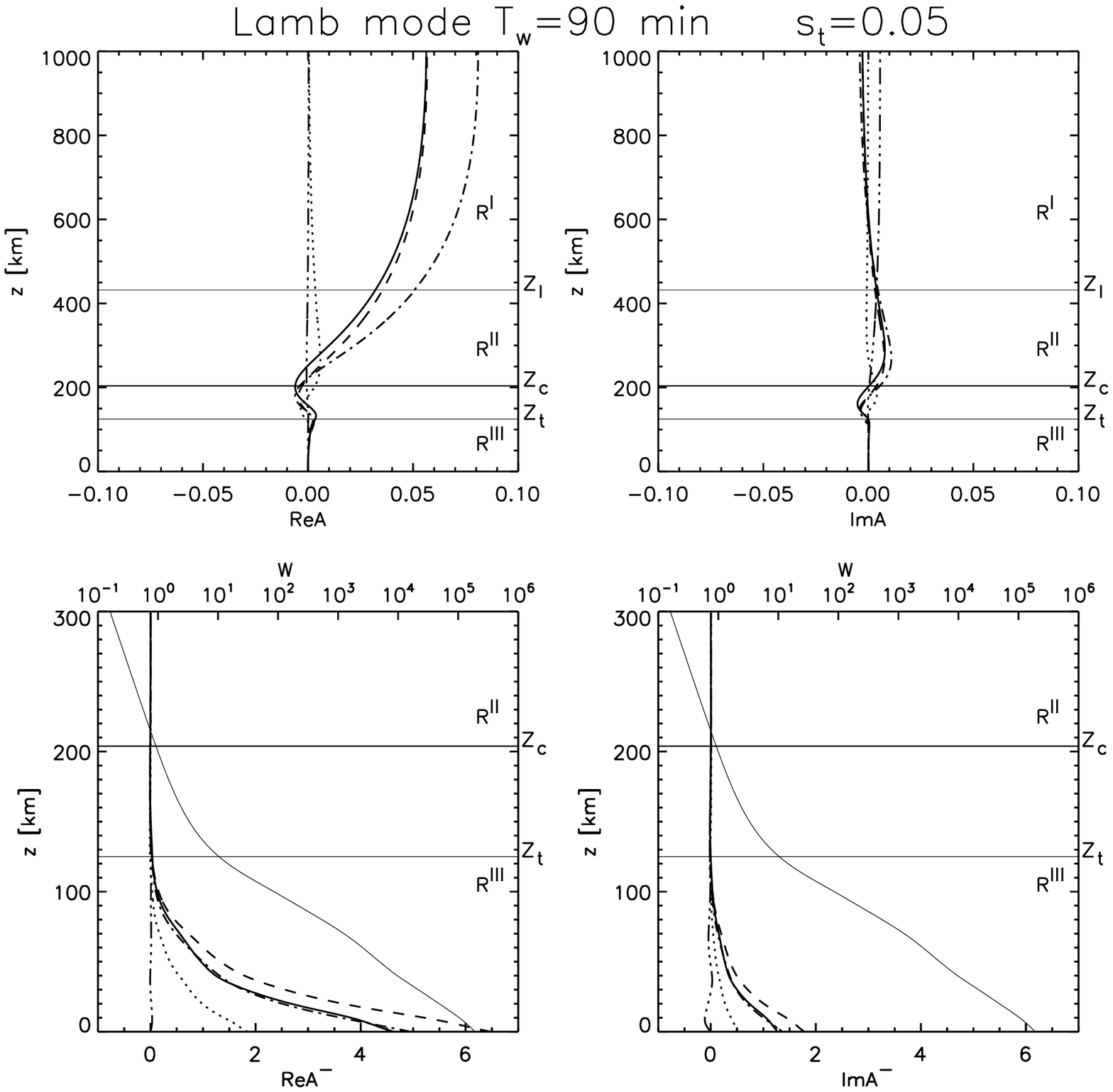}\\
  \caption{The Lamb wave with the period  $T_w=90 min$.}\label{fig18}
\end{figure}

\begin{figure}
    \includegraphics[width=1\linewidth]{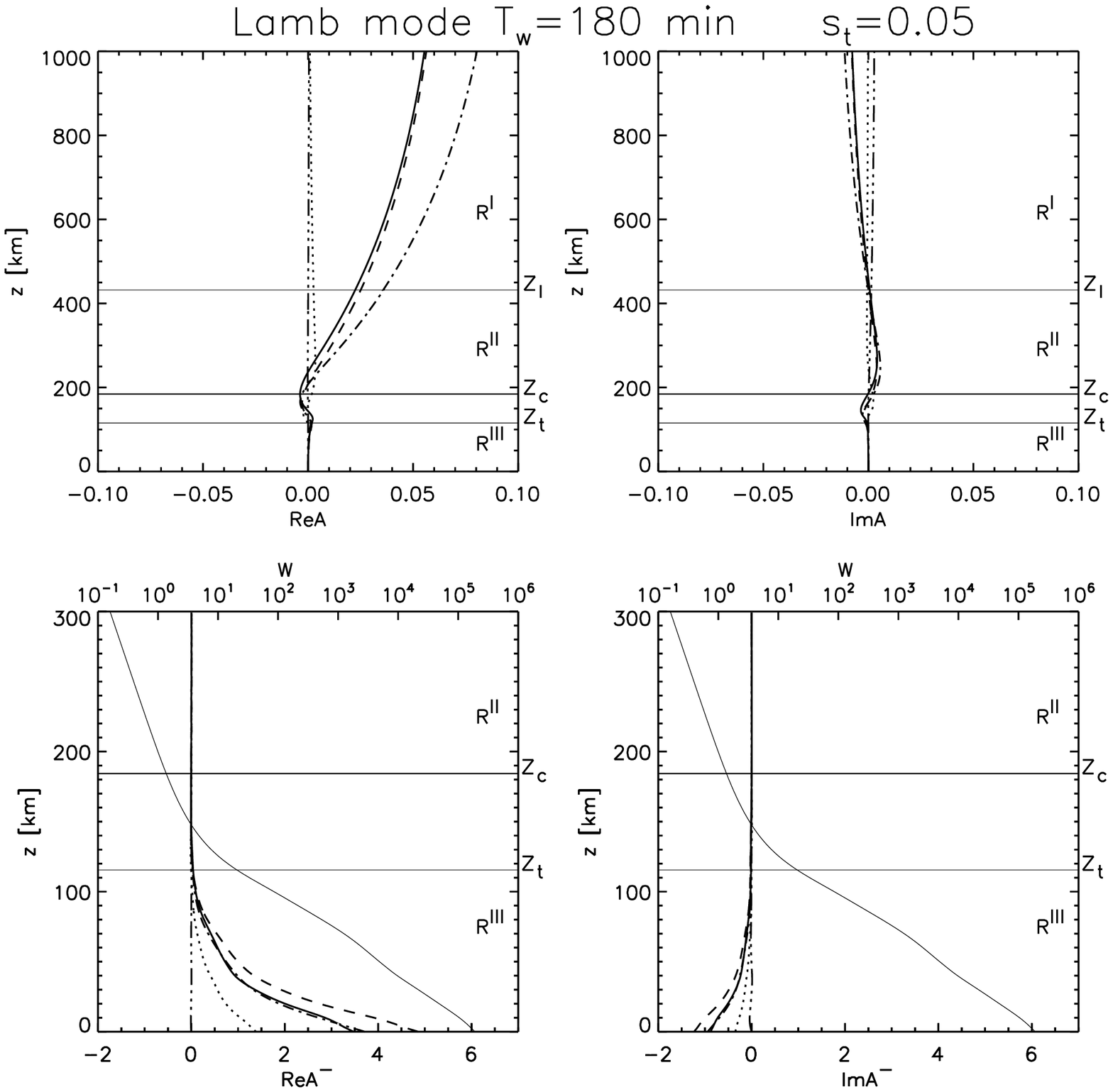}\\
  \caption{The Lamb wave with the period  $T_w=180 min$.}\label{fig19}
\end{figure}

\begin{figure}
    \includegraphics[width=1\linewidth]{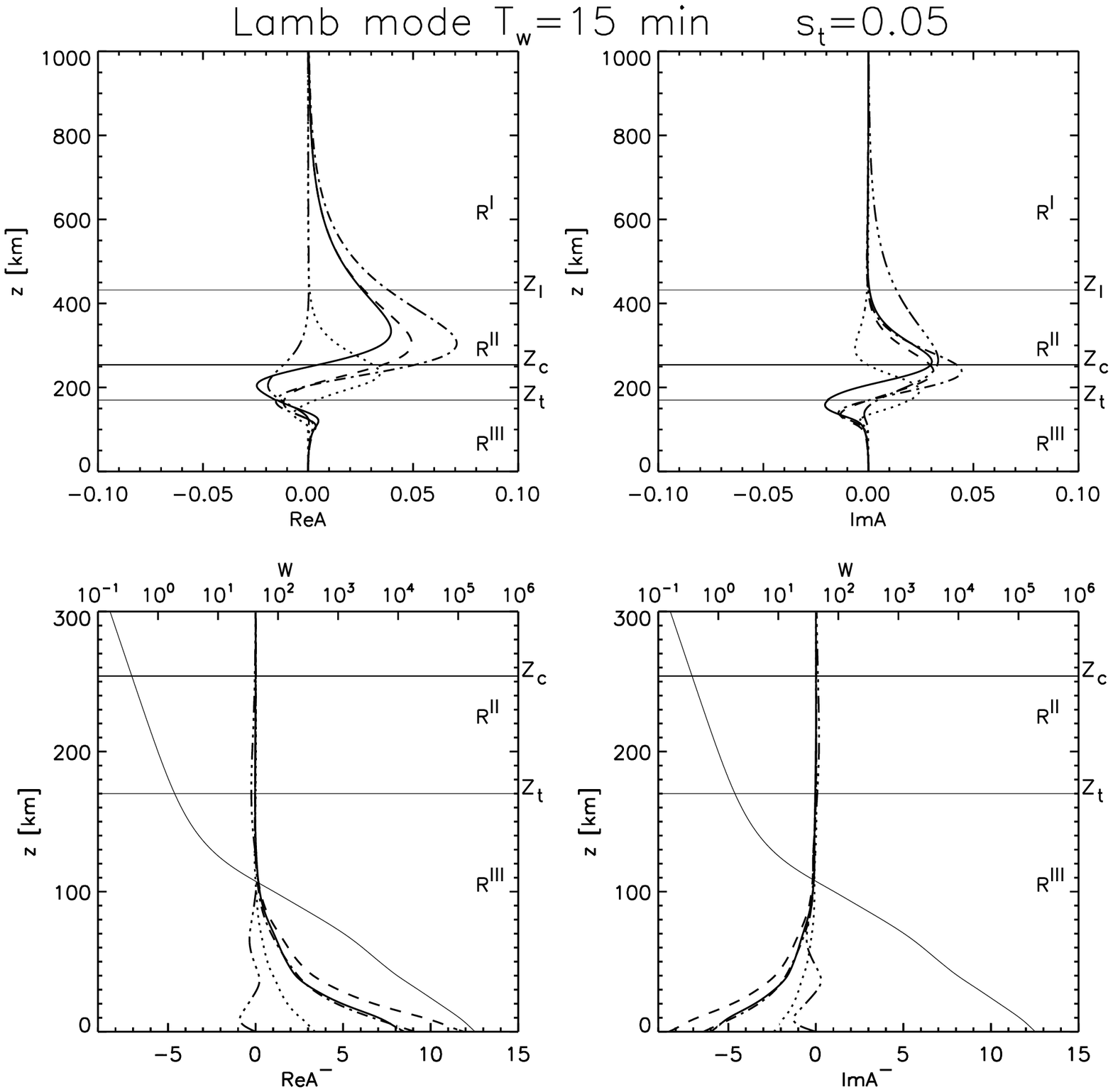}\\
  \caption{The Lamb wave with the period  $T_w=15 min$.}\label{fig20}
\end{figure}

\begin{figure}
    \includegraphics[width=1\linewidth]{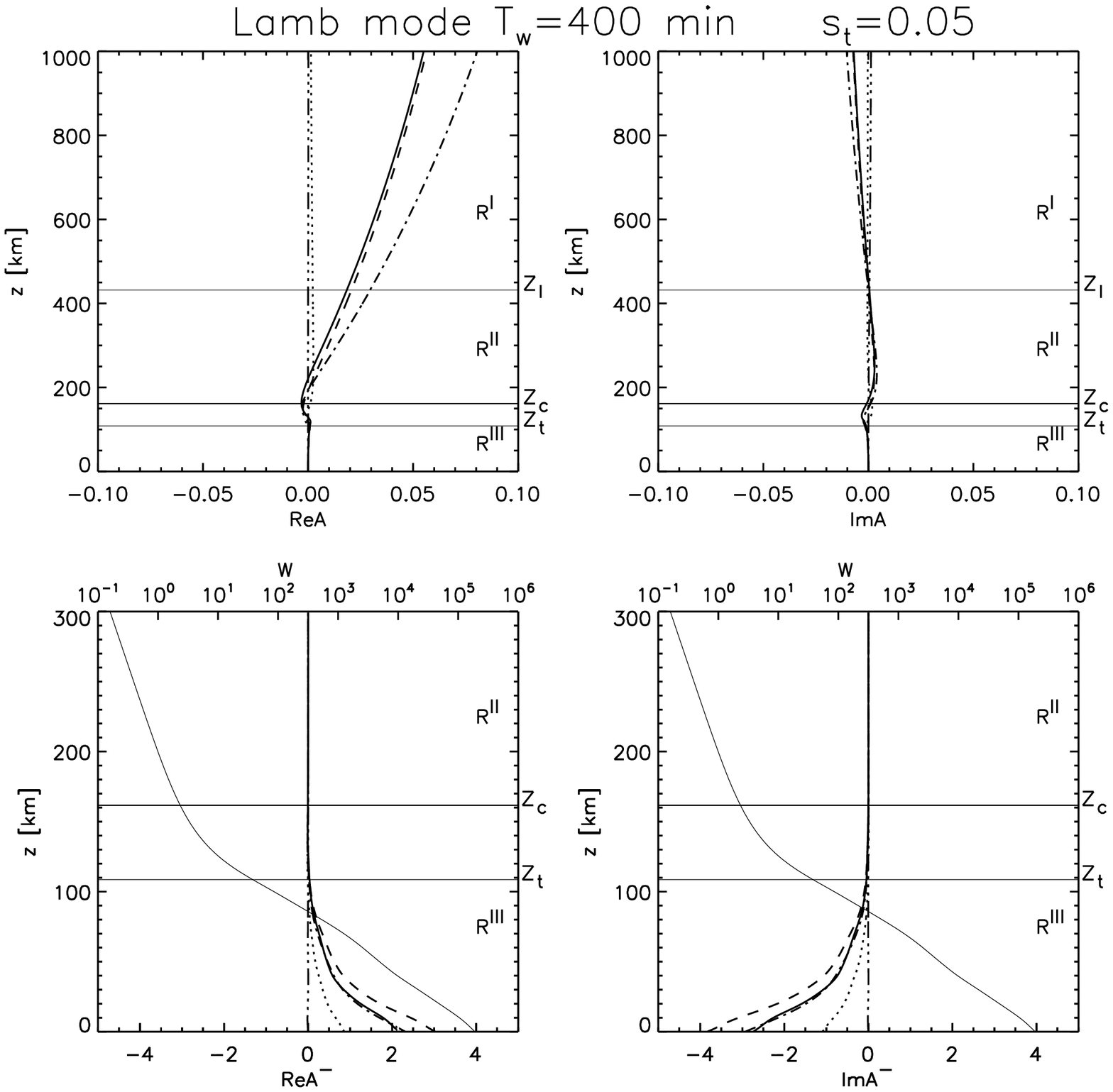}\\
  \caption{The Lamb wave with the period  $T_w=400 min$.}\label{fig21}
\end{figure}

The IGW waveguide modes are not the only of the trapped
atmospheric low frequency waves. Surface Lamb waves can be called
trapped waves. A domination of pressure, horizontal propagation
velocity close to the value of sound velocity near to the Earth's
surface and extremely weak horizontal attenuation \cite{Hook} are
typical of these waves. The DSAS method has turned to be suitable
for description of such  modes. The results obtained for Lamb
waves are presented in Figures \ref{fig17}-\ref{fig21}.
We do not give the dependence between the horizontal attenuation
parameter and period. Predictably, a ratio $\frac{{\rm
Im}k_x}{{\rm Re}k_x}\approx  1.5\times  10^{-5}$ is extremely
small with little variations throughout the spectrum. Figures
\ref{fig18}-\ref{fig21} give a presentation of the wave structure
of a Lamb mode, the same as in Figure \ref{fig13},
\ref{fig14}-\ref{fig16}. From the last figures, we see clear
domination of a pressure component (dotted lines.) Besides, we see
that, despite exponential attenuation of a mode in the lower
atmosphere, the upper atmospheric amplitudes of relative values
strongly rise, on the contrary, relative to the lower atmosphere.
Thus, this mode can be present at the ionospheric heights.

It is also possible to satisfy the conditions of trap for the
infrasonic waves higher  the sound cutoff frequency. Dispersion of
such modes has been analyzed in the WKB-approximation (see, for
example, \cite{Pon}) Unlike IGWs, the infrasonic waveguides have a
multimode character and divides in their localization place in the
lower or middle atmosphere. The waveguide signals of such a type
are also widely represented in real disturbances in the lower and
middle atmosphere. Analysis of such modes based on the solutions
above a source can be also carried out and is quite possible as a
subject of our method development; however, these modes deserve
separate consideration because of the variety of their
characteristics and features.
\section{Conclusion}\label{section5}
We have studied low frequency trapped waves such as IGW waveguide
modes and surface Lamb waves based on the DSAS method.

We have focused on the waveguide solutions corresponding to
trapped waves like IGWs. The so-called wave leakage to the upper
atmosphere is typical of such trapped waves. Because of an
exponential factor typical of the atmospheric waves, upward
disturbance exponentially grows in its relative values. The latter
leads to the fact that trapped waves are strongly manifested at
large heights causing "visible" travelling ionospheric
disturbances (TIDs) to large distances. For the model of the
atmosphere that we chose, we have obtained that we had only one
fundamental (nodeless) mode at the set frequency. We have
calculated a dispersion curve for an IGW waveguide, a complex
horizontal number curve as a frequency function. For the first
time, we have fully described a vertical structure of an IGW mode
at all the heights together with its part leaking from a waveguide
and fully accounting for dissipation. We have obtained such a
possibility thanks to use of the method for the DSAS
\citep{Dmitrienko2016} construction. This paper has also revealed
that a dissipationless description turned sufficiently useful in
the lower atmosphere. It allows us to estimate the number of modes
and obtain testing values for a numerical calculation of the
eigenvalues. The WKB method is the most convenient for these
purposes. It is interesting to note that the WKB description gives
correct results, despite its formal inapplicability, with a good
coincidence of the WKB dispersion dependences with the exact ones.

Using the obtained dispersion and wave propagation characteristics
of the leakage disturbance, we have obtained a very good match
with the major characteristics of the observed TIDs: a correlation
of horizontal scales with wave periods, propagation ability to
many thousands of kilometers without significant attenuation;
reverse direction of vertical phase velocity; small values of
vertical phase velocity; specific inclination of the phase front.

The dispersive characteristics and complete spatial description of
vertical wave structure have been also obtained for Lamb waves.

We believe that a possibility of calculating an adequate vertical
structure in case of trapped waves have to give a possibility of
tracing atmospheric disturbances at its high frequencies,
including, in the ionosphere, with long distance propagation of
these disturbances.

 \textbf{Acknowledgment}
We are grateful to Dr A.V. Medvedev and Dr K.G. Ratovsky  for
helpful discussion in the course of writing this paper.
\bibliographystyle{elsarticle-harv}

\end{document}